# PAPR Reduction of OFDM System Through Iterative Selection of Input Sequences

Md. Sakir Hossain[1], Sabbir Ahmed[2], Enayet Ullah[3], Md. Atiqul Islam[1]

[1]International Islamic University Chittagong, Bangladesh
[2]Ritsumeikan University, Japan
[3]University of Rajshahi, Bangladesh
shakir.rajbd@yahoo.com

**Abstract:** Orthogonal Frequency Division Multiplexing (OFDM) based multi-carrier systems can support high data rate wireless transmission without the requirement of any extensive equalization and yet offer excellent immunity against fading and inter-symbol interference. But one of the major drawbacks of these systems is the large Peak-to-Average Power Ratio (PAPR) of the transmit signal which renders a straightforward implementation costly and inefficient. In this paper, a new PAPR reduction scheme is introduced where a number of sequences from the original data sequence is generated by changing the position of each symbol and the sequence with lowest PAPR is selected for transmission. A comparison of performance of this proposed technique with an existing PAPR reduction scheme, i.e., the Selective Mapping (SLM) is performed. It is shown that considerable reduction in PAPR along with higher throughput can be achieved at the expense of some additional computational complexity.

## I. INTRODUCTION

OFDM is a very attractive technique for high speed data transmission in mobile communications due to various advantages such as high spectral efficiency, robustness to channel fading, immunity to impulse interference, and capability of handling very strong multipath fading and frequency selective fading without having to provide powerful channel equalization. Considering these, it has already been adopted as the standard transmission technique in the wireless LAN systems [1],[2] and the terrestrial digital broadcasting system [3]. It is also being considered as one of the candidate transmission techniques for the next generation of mobile communications systems. One drawback of OFDM technique is the larger peak to average power ratio (PAPR) of its time domain signal [4], [5]. The problem stems from the fact that high PAPR can drive transmitter power amplifiers in to saturation region causing inter-modulation noise and there-by inflict severe degradation of bit error rate (BER) performance. The simplest solution to overcome this problem is to operate the non-linear amplifier at the linear region by taking out enough power from the input, i.e., input back-off. However, this method degrades power efficiency of the overall system and hence is not suitable for power constrained handheld devices like mobile terminal and portable wireless LAN terminal etc. Thus implementation of OFDM technique in a power efficient manner has drawn significant research attention in the recent past and many methods have been proposed with the objective of PAPR reduction

A survey of PAPR reduction techniques reveals that perhaps the most widely known methods are signal clipping [6], block coding [7-8], selected mapping SLM [9] and partial transmit sequence (PTS) [10-12]. Of them, clipping is a very simple method to reduce PAPR . This can lower the PAPR easily by cutting away the signal above the assigned clip level. But it results in out-of-band radiation and in-band distortion causing poor signal quality. Block coding method can reduce PAPR significantly. It has been used in Magic WAND (wireless ATM network demonstrator) system because it does not degrade the OFDM signal and shows an additional coding effect. However, the code rate and bandwidth efficiency are very low. Also, computation is exponentially increased with the number of subcarriers. Finally, SLM and PTS are the phase control method to reduce PAPR. SLM multiplies an OFDM data by several phase sequences in parallel and selects the data sequence of the lowest PAPR among them. PTS divides the input OFDM data into several clusters and phase rotation factors (or combining sequences) are multiplied to get the low PAPR signal. Although these two methods can reduce PAPR effectively without any signal distortion, the side information about the phase rotation must be transmitted to the receiver. Error in the side information can cause significant BER performance degradation. Furthermore, system complexity considerably goes up because of many IFFT (inverse fast Fourier transform) stages and the long phase optimization processes.

In this paper, we propose a new probabilistic scheme to reduce PAPR based on the selection of input sequence. This proposed technique produces all possible permutation of input sequence and the sequence with minimum PAPR is transmitted. The simulation result show that the proposed method can show better PAPR reduction performance than SLM with relatively less side information.

## II. OFDM SYSTEM MODEL

Let us define N symbols in OFDM are { $X_n$, n=0,1 2,...N-1}. A set of N sub-carriers, i.e., { $f_n$, n=0,1,2....,N-1}, is used for these symbols in the OFDM. The N sub-carriers are chosen to be orthogonal, that is, $f_n=n\Delta f$ in frequency domain, where $\Delta f=1/NT$ and T is the OFDM symbol duration. The basic OFDM transmitter and receiver is are shown in figure 1. The OFDM signal is expressed as

$$x(t) = \frac{1}{\sqrt{N}} \sum_{n=0}^{N-1} X_n e^{j2\pi f_n t}, \quad 0 \le t \le T \quad (2.1)$$

In practical systems, a guard interval (cyclic prefix) is inserted by the transmitter in order to remove intersymbol interference and interchannel interference (ICI) in the multipath environment. However, it can be ignored since it does not affect the PAPR.

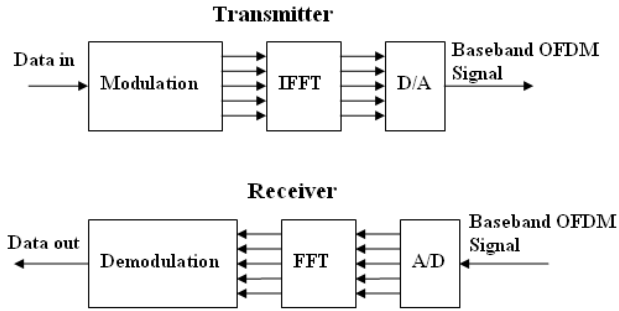

Figure 1: Basic OFDM Transmitter and Receiver

The PAPR of the transmit signal s(t), defined above, is the ratio of the maximum instantaneous power and the average power, given by

$$PAPR = \frac{\max |x(t)|^2}{E\left[|x(t)|^2\right]} \quad (2.2)$$

where E{.} denotes expectation. In some frames of OFDM signals, large PAPR happens since the structure of the given symbols may cause these peaks. IFFT operation can be viewed as multiplying sinusoidal functions to the input sequence, summing, and sampling the results. Thus the high correlation property of IFFT input causes the sinusoidal functions to be arranged with in-phase form. After summing the in-phase functions, the output might have large amplitude.

Moreover, with the increase of number of subcarriers, PAPR of the resulting system also increases. The reason for this is that when the number of subcarriers is large and they all are added in some positive or negative phases, the resulting amplitude becomes large enough to exceed saturation point of high power amplifier (HPA). Figure 2 shows such situation.

## III.   PAPR REDUCTION STRATEGIES

A number of methods are available for PAPR reduction. But each of these methods provides some advantages from others creating some problems also. Here we introduce a new PAPR reduction strategy that is based on iterative selection of input sequences. In this technique, different permutations of input data are made and the permutation with the least PAPR is selected for transmission. But in SLM, the input sequence is modified by multiplying it with number of  phase vectors and the sequence with the minimum PAPR  is selected for transmission. Another advantage of proposed technique in addition to the least PAPR is that it does not need any multiplication. Moreover, the state space for finding minimum PAPR is also considerably large. However, in this section Selective Mapping (SLM) and our proposed technique will be described in detail.

### A.   SLM Method

A block diagram of Selected Mapping (SLM) is shown in Figure 3. In the SLM, the U-1 statistically independent phase sequences are generated. Symbol sequences are multiplied by the U-1 different phase sequences whose length is equal to the number of carriers before the IFFT process. The PAPR is calculated for the U-1 phase rotated symbol sequences and one original sequence. Then the symbol sequence with the lowest PAPR is selected and transmitted. The used phase sequence is transmitted as the side information. The receiver performs the reverse rotation to recover the data symbol. The SLM needs the IFFT process whose number is equal to U and thus a large amount of calculation.

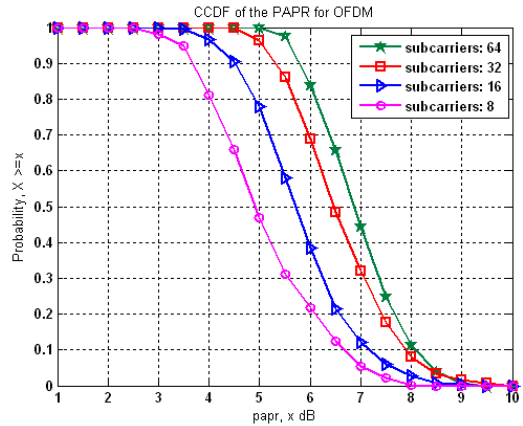

Figure 2: Illustration of the effect of number of subcarriers on PAPR when FFT size is fixed at 128.

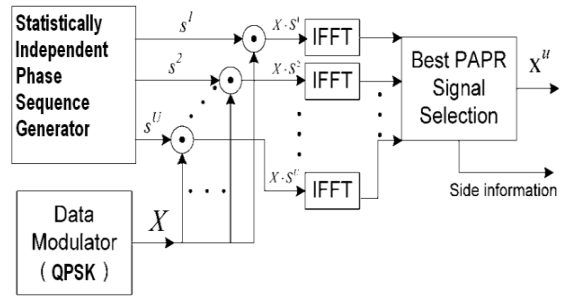

Figure 3:  OFDM transmitter model for the SLM approach.

### B.   PROPOSED METHOD

Here we propose a new statistical scheme, which we shall call in this paper Iterative Selection of Input Sequence (ISIS), which can reduce PAPR to a minimal level. The block diagram of the proposed system is shown in figure 4.

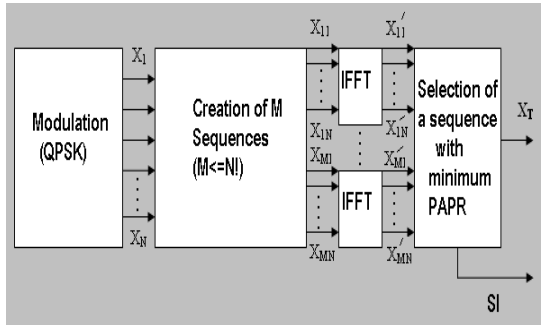

Figure 4: Block diagram of ISIS approach

The proposed method can be explained as follows:

*Step-1: Take the input sequence to be transmitted to a buffer.*

*Step-2: Produce all possible permutations of the buffered sequence. If the number of elements in the input sequence is N then the maximum number of permutations would be N!.*

*Step-3: Pass each parallel sequence through IFFT separately and compute the PAPR for each sequence. Put the resulting PAPR in buffer.*

*Step-4: Finally find the sequence for which PAPR is minimum and transmit that sequence.*

*Step-5: As side information, send the serial number of the transmitted sequence in the permutation table through a secure channel or it may be transmitted number of times and at the receiver majority voting technique may be applied for finding the correct sequence number.*

*Step-6: At the receiver, produce all possible permutation of the received sequence and place them in buffer.*

*Step-7: Produce permutations of each buffered sequence and compare the SI-th sequence to the received sequence. If they are all the same. Then the current sequence whose permutation is created is the actual desired sequence that was altered and transmitted.*

An example of this method would be: For example, if original data sequence is [A.B,C,D], then there are 24 combinations of these four symbols and say, the sequence [B,C,D,A] produces a minimum PAPR. The sequence [B,C,D,A] is selected for transmission. If the index of this sequence in the permutation table is S then S is sent as side information. At the receiver, permutation of [B,C,D,A] is taken and the permutation of each sequence obtained from permutation of [B,C,D,A] is taken. Then match the received sequence with S-th sequence of each 24 set of sequence. The sequence for whose S-th sequence is the received sequence (that is, [B,C,D,A]) is the original data sequence for which [B,C,D,A] was sent from transmitter.

However, input sequence can also be changed on binary data and recovered in the receiver in binary form.

## IV. SIMULATION RESULTS

Here we simulate figure 3 and figure 4 using Matlab and compare the results. Also we try to enhance the performance of the ISIS approach. The simulation parameters are given in Table I.

Table 1: Simulation Parameters

| Modulation Method | QPSK |
|---|---|
| The number of FFT points | 16 |
| The number of subscribers | 8 |
| No. of frame transmitted | 512 |

It is found from figure 5 that SLM reduces PAPR significantly. Figure 5 also compares the performance of Walsh-Hadamard sequence and Golay complementary sequence as techniques for producing phase rotation vector for SLM method. The performances of both sequences are almost same. An important thing that is found from this figure is that a considerable amount of more PAPR reduction is obtained by the proposed technique than the SLM. It is seen that PAPR of proposed technique is about 2 dB lower than SLM. The reason for this result can be explained as: IFFT actually multiplies input sequences by sinusoidal functions, sum different signals after multiplication, take sample after summation. If we rearrange the input sequence number of times and perform multiplications, summation and sampling for each arrangement, then maximum peaks for each set of samples will be different. If we choose the time domain signal with least maximum peak, then signal with minimum PAPR can be achieved. For maximum number of rearrangement minimum PAPR can be achieved. If OFDM frame size is n then maximum number of arrangement that can be achieved is n!. Thus to get minimum PAPR, n! arrangements are needed to be used. Thus the minimum PAPR can be achieved from this technique but with relatively more complexity.

It is found from Figure-5 that for minimum PAPR, the number of sequences that are needed to use is n!. The effect of number of sequences on PAPR (i.e., how does PAPR vary with number of sequences used) is investigated below. The effect of total number of sequences used on proposed method's performance is shown in Figure-6.

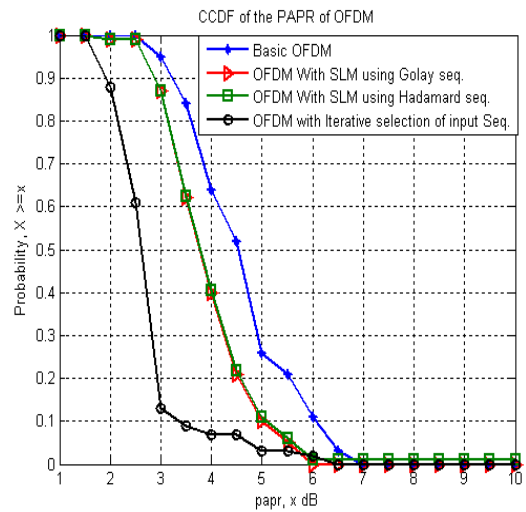

Figure 5: Comparison between SLM with two sequences and proposed ISIS method for PAPR reduction.

It is seen from figure-6 that with the increase of number of sequences used, PAPR reduces. It is seen that PAPR performance is same as SLM when number of sequences used is 8 A considerable reduction of PAPR is achieved when 100 sequences are used. Using 500 sequences gives nearly same result as n! number of sequences. But using 1000 sequences shows the same result as n!. Thus about 40 times complexity can be reduced using smaller number of sequences without any degradation of PAPR performance.

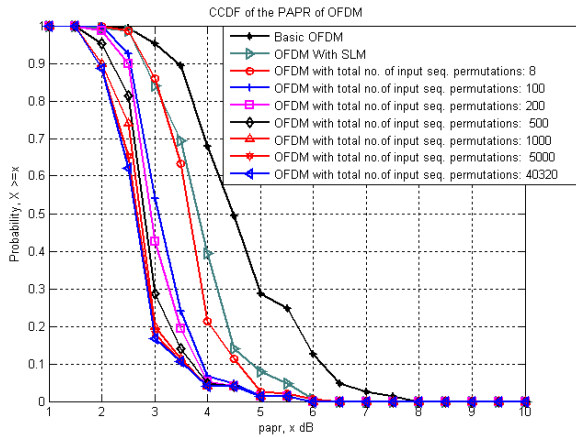

Figure 6: Effect of number of total number of sequences used on PAPR reduction

Figure 7 shows the effect of OFDM frame size on the performance of proposed technique. It is seen from figure 7 that for basic OFDM, PAPR increases for greater frame size (that is, for greater number of subcarrier). It is also seen that the proposed technique performs better for larger frame size. The reason for this is that as frame size increases the number of possible combinations also increases. Hence the probability to find signal with lower PAPR also rises.

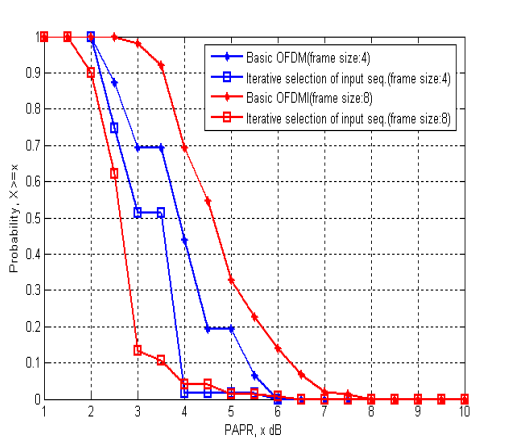

Figure 7: Effect of number of subcarrier on the performance of proposed technique

## V.  CONCLUSION

In this paper, we proposed a PAPR reduction scheme and compared its performance with the well known SLM technique. We showed that with proposed method, significant PAPR reduction can be achieved compared to SLM. In addition, the requirement of side information is also lesser here, i.e., log2 (no number of phase rotation vector) in SLM and only a single symbol in our scheme. Moreover, in SLM the number of phase rotation vectors is needed to be increased for better PAPR reduction. The main overhead of our system is the computation complexity which goes up with the number of possible input sequences. But even with a small number of chosen input sequences, our system shows similar PAPR reduction capability compared to SLM with much less computation overhead since there is no multiplication involved as is the case with SLM.